\documentclass[12pt,draftclsnofoot,onecolumn]{IEEEtran}
\usepackage{graphicx,cite,epsfig,amssymb,amsmath,txfonts,bm,
subfigure,url,stfloats,latexsym,booktabs}

\begin{document}
\markboth{International Journal of Communication Systems, Vol. XX,
No. Y, Month 2010}{}

\title{\mbox{}\vspace{1.5cm}\\
\textsc{Data Fusion Based Interference Matrix Generation for
Cellular System Frequency Planning}
\vspace{1.5cm}}

\author{\normalsize
Zhouyun Wu, Aiping Huang{$^{^\dagger}$}, Haojie Zhou, Cunqing Hua and \mbox{Jun
Qian}
\thanks{Zhouyun~Wu (e-mail: {\tt wuzhouyunjing@zju.edu.cn}), Aiping~Huang (e-mail: {\tt aiping.huang@zju.edu.cn}), Haojie~Zhou (e-mail: {\tt xingyu.boy@gmail.com}) and Cunqing~Hua (e-mail: {\tt cqhua@zju.edu.cn}) are with the Institute of Information and Communication
Engineering, Zhejiang University, and Zhejiang Provincial Key
Laboratory of Information Network Technology, China. Jun~Qian
(e-mail: {\tt qianjun@zj.chinamobile.com}) is with China Mobile Group Zhejiang Company Limited, Hangzhou, China.
}
\thanks{The paper was submitted 7 Octorber 2010.}\\
\vspace{1.5cm}
\underline{{$^{^\dagger}$}Corresponding Author's Address:}\\
$\mbox{Aiping~Huang}$\\
Institute of Information and Communication Engineering\\
Zhejiang University\\
Zheda Road, Hangzhou, 310027 China\\
Tel: +86-571-87951709\\
Fax: +86-571-87952367\\
E-mail: {\tt Aiping.huang@zju.edu.cn}}

\date{\today}
\renewcommand{\baselinestretch}{1.2}
\thispagestyle{empty} \maketitle \thispagestyle{empty}
\newpage
\setcounter{page}{1}

\begin{abstract}
Interference matrix (IM) has been widely used in frequency planning/optimization of cellular systems because it describes the interaction between any two cells. IM is generated from the source data gathered from the cellular system, either mobile measurement reports (MMRs) or drive test (DT) records. IM accuracy is not satisfactory since neither MMRs nor DT records contain complete information on interference and traffic distribution. In this paper, two IM generation algorithms based on source data fusion are proposed. Data fusion in one algorithm is to reinforce MMRs data, using the frequency-domain information of DT data from the same region. Data fusion in another algorithm is to reshape DT data, using the traffic distribution information extracted from MMRs from the same region. The fused data contains more complete information so that more accurate IM can be obtained. Simulation results have validated this conclusion.
\end{abstract}

\begin{keywords}
\centering
frequency plan, interference matrix, data fusion, mobile measurement reports, drive test records
\end{keywords}
\newpage

\IEEEpeerreviewmaketitle

\section{INTRODUCTION}

Frequency planning is of great importance to the large-scale cellular systems. A set of frequencies is allocated among thousands of cells/cell-clusters in order to satisfy the capacity demand of each cell. Limited frequencies are reused over non-overlapped space and the inter-cell interference is avoided owing to the frequency separation. In contrast to dynamic frequency allocation for small-scale wireless networks or cognitive wireless networks, a frequency plan generated using a centralized algorithm will be employed several months in a large-scale cellular system until there is an obvious change of wireless environment or capacity demand.

Frequency planning for large-scale cellular systems has been a hot research issue for decades. Early research focused on the frequency assignment to minimize the number of frequencies needed while satisfy capacity demand and frequency separation constraint [1-12]. Here frequency assignment is formulated as an optimization problem, and the constraint of frequency separation between cell pairs is characterized by a compatibility matrix (also known as channel separation matrix) or exclusion matrix.

In recent years, another type of practical frequency planning/optimization has drawn more attentions. It aims to minimize the intra-system interference through making full use of the available frequency band owned by the network operator [13-17] under the condition of actual wireless environment. Interference matrix (IM) is employed to describe the interaction between any two cells in a cellular system. An element of IM is an indicator of the potential interference intensity between the two cells assuming that the two cells operate at the same frequency. The element value depends on the distance and propagation condition between the two cells, the base station parameters and traffic distribution of the cells.

The accuracy of an IM mainly depends on its source data. Source data comes from one of the two categories: model-generated and system measurement. The models used to generate source data include propagation model \cite{c13, c14} such as Okumura-Hata and ray-tracing models \cite{c15}. These model-generated data do not reflect the real interaction between practical cells, thus resulting in impractical IM. System measurements are obtained during the operation of actual cellular system. The two most commonly used system measurements are (1) mobile measurement reports (MMRs) sent by active user terminals to the base transceiver station (BTS) in a dedicated mode [18-20], and (2) drive test arranged for routine system optimization \cite{c16, c17}.

MMRs were originally designed for seamless handover between cells. Mobile phones send MMRs to the BTS every 480ms, which contain 7 signal strength measurements from the serving cell and 6 neighboring cells respectively. The received signal strengths from these 6 neighboring cells are the strongest among those of all the cells in BA (BCCH Allocation) list. For simplicity of presentation, these 6 neighboring cells are called "strongest neighboring cells" in the following. A BA list contains limited number of neighboring cells of the serving cell, such as 32 for GSM, and is preset manually by network operator. Being used in frequency planning/optimization, these signal strength measurements are transformed to MMRs data reflecting the wireless environment that user experienced, and the number of MMRs is a measure of traffic. As a result, IM generated from MMRs gives larger weight to the region with heavier traffic, and the follow-up frequency planning/optimization can pay more attentions to the region. However, some of the neighboring cell signals might be omitted from MMRs due to outdated BA list or undistinguishable cells operating at the same frequency. Although it was concluded through simulations that this problem would not degrade the accuracy of the MMRs based IM evidently \cite{c21, c22}, MMRs' incompletion of frequency-domain information is severe in those countries and regions where the cellular network is adjusted frequently due to network's fast expanding, traffic's explosive increase, GSM900 and DCS1800 co-existence, and transition from 2G to 3G. This is caused by several reasons: (1)  BA list update is not in time; (2) part of the 6 measurements might wrongly report the signal strengths of other frequency bands (such as DCS1800) or other systems (such as TD-SCDMA or WCDMA), so that less than 6 measurements are available for the system itself (such as GSM900). Some of the elements in MMRs based IM might be wrongly set to zero as a result of incomplete information on the potential interfering neighboring cells.

Drive test (DT) is done by the frequency sweeper carried on a vehicle traveling along roads, for the purpose of obtaining good knowledge of the radio environment. DT records are of high accuracy and consist of measurements of all the frequencies over large frequency band. The number and distribution of DT records depend on the vehicle velocity and road situation, thus are uselessless to frequency planning/optimization. As a source data of IM, DT data transformed from DT records offers complete information of potential interfering neighboring cells, but it can not reveal the actual communication traffic distribution.

In this paper, two IM generation algorithms based on source data fusion are proposed. The fused data contains more complete information so that more accurate IM can be obtained. The steps of these two algorithms are described as follow. First, DT data of the serving cell is clustered into $K$ clusters according to their frequency-domain information, each cluster has unique characteristic and corresponds to a specific geographic region in the cell. Secondly, the number of MMRs in each cluster is calculated through a multiple linear regression, which is the estimation of the traffic amount generated in the corresponding geographic region. Thirdly, a data fusion is carried out to reinforce MMRs data, using the frequency-domain information of DT data from the same region; or another data fusion is performed to reshape DT data, using the traffic distribution information extracted from MMRs from the same region. Finally, the reinforced MMRs data or the reshaped DT data is used to calculate IM. These two proposed algorithms are named as MMRs+DT algorithm and as DT+MMRs algorithm, respectively. They have the same steps of DT data clustering, MMRs classifying and traffic distribution estimation as well as IM calculation, but different data fusions.

The rest of this paper is organized as follows. In Section II, source data and conventional IM generation algorithm are introduced. The proposed MMRs+DT algorithm and DT+MMRs algorithm are described in Sections III and IV, respectively. Simulation results are given in Section V to verify the improvement in the information completeness by utilizing two source data. The main conclusions are drawn in Section VI.

\vspace{0.2in}
\section{SOURCE DATA AND CONVENTIONAL IM GENERATION ALGORITHM}\label{sec:data and algorithm}

For simplicity of presentation and symbolization, the following discussion will focus on the scenario of one serving cell with its neighboring cells. It can be easily generalized to normal cellular system scenario.

\subsection{MMRs Data}\label{subsec:MMRs Data}

Assume that $J$ neighboring cells' signal strength measurements are reported in MMRs. From the signal strength measurements reported in MMRs, the carrier-to-interference ratio (CIR) of serving cell $u$ and neighboring cell $v$ can be calculated. Assume there are $Q$ successive and non-overlapping intervals of CIR value, with interval index \mbox{${q = 1, \ldots ,Q}$}.

MMRs data is defined as the number of CIR values falling into each particular CIR interval, and is expressed as a  $JQ$-dimensional column vector \mbox{${\bf{R}} = [{\bf{R}}_1^{\rm{T}} {\kern 1pt}
{\bf{R}}_2^{\rm{T}} {\kern 1pt} \ldots {\bf{R}}_j^{\rm{T}} {\kern
1pt}  \ldots {\bf{R}}_J^{\rm{T}} {\kern 1pt} ]^{\rm{T}}$}. Where each $Q$-dimensional column vector  \mbox{${{\bf{R}}_j}$} is the $j$th column in Figure 1. The element in the [\mbox{${ \left( {j - 1}
\right)Q + q }$}]th column of \mbox{$ \bf{R} $} is denoted as \mbox{$r_{j,q} $} , which is the number of CIR values against neighboring cell $j$ falling into CIR interval $q$. For example, $r_{3,2}$ is the number of CIR against neighboring cell 3 falling into CIR interval 2, and $r_{3,2}  = 100$ in Figure 1.

\subsection{DT Data}\label{subsec:DT data}

Assume that there are $M$ DT records containing signal strengths from $I$ neighboring cells. DT data is expressed as a matrix \mbox{$ {\bf{D}} = [{\bf{d}}_1{\kern 1pt} {\bf{d}}_2  \ldots {\bf{d}}_m  \ldots {\bf{d}}_M ] $} where the $m$th column vector \mbox{$ {\bf{d}}_m $}  consists of the CIR values against all the $I$ neighboring cells while the CIR values are calculated from the signal strengths in the $m$th DT record. That is, the element \mbox{$ d_{i,m} $}  in the $i$th row and $m$th column of ${\bf{D}}$ is the CIR value against neighboring cell $i$, which is calculated from the signal strengths in the $m$th DT record.

$I$ and $J$ are usually not equal since both measurement methods and tools of getting MMRs and DT records are different.

\subsection{Generating IM}\label{subsec:Generating IM}

As a form of IM, inter cell dependency matrix (ICDM) is widely used owing to its accuracy and simplicity [18].
Each element of ICDM is an estimation of the probability of a CIR value being lower than a preset threshold \mbox{$ CIR_{{\rm{threshold}}} $} , or equivalently, the probability of a CIR interval index being lower than a preset threshold $Q_{{\rm{threshold}}}$, $Q_{{\rm{threshold}}}  < Q$. The procedure of obtaining ICDM element generated from MMRs data and DT data are shown in Figure 2 and Figure 3, respectively.

\vspace{0.2in}
\section{IM GENERATION ALGORITHM BASED ON MMRS DATA REINFORCED BY DT DATA}\label{sec:MR+DT}

The MMRs data and DT data to be fused should be from the same geographical position or region, since only they are the descriptions of the wireless environment there. Therefore, the MMRs data and DT data from the same position or region have to be identified and separated from the whole MMRs data and DT data first.

Frequency spectrum consists of the signal strengths of all the signals from all the cells. The relative strengths at different frequencies correspond to the position where the MMRs and/or DT records were measured. The CIR values against all the neighboring cells can be obtained from the frequency spectrum, and then constitute a "CIR spectrum". CIR spectrum (CIRS) can be regarded as the "fingerprint" of a position as it is closely related to the position where the wireless signals are measured. CIRS can be derived from DT records, since DT records have complete frequency-domain information as addressed in Section I.

MMRs data is the numbers of CIR values against all the $J$ neighboring cells falling into different CIR intervals. Therefore, it is essentially a statistic of CIRS at different positions. According to the characteristic of CIRS, MMRs data from different positions can be identified and classified by regression analysis, provided that all the CIRSs are known.

To reduce the number of CIRS types in the serving cell and thus reduce the computational complexity of regression analysis, all the $M$ CIRSs derived from DT data are clustered into $K$ clusters by clustering analysis, and each cluster is related with a specific region in the serving cell. Therefore, MMRs data will be identified and classified by multiple linear regression on the basis of cluster.

The block diagram of this algorithm is shown in Figure 4. The CIRS profile, clustering, regression and data fusion will be discussed in this section in turn.

\subsection{CIRSP and SP Matrix}

CIRS derived from DT data is the "fingerprint" of geographical position. Note that the main characteristic of CIRS is from the 6 neighboring cells with the strongest signals, i.e., from the so called "the 6 strongest neighboring cells". For data fusion, CIRS profile (CIRSP) is defined here. It is transformed from CIRS by reserving the information of the 6 strongest neighboring cells and removing those of all the other cells, thus is much simpler than CIRS. The CIRSP element corresponding to a neighboring cell is set to q, if the neighboring cell is one of the 6 strongest neighboring cells and the CIR against this neighboring cell falls into the $q$th CIR interval. The CIRSP elements corresponding to the other neighboring cells are set to zero. The CIR intervals mentioned here agree with those in MMRs data calculation in Section II-A.

An example of the CIRSP is shown in Figure 5. The serving cell is in the center and the serial numbers of its neighboring cells are marked. The CIRSPs from locations A and B are plotted on the right of the figure. It can be observed that location A is close to the center of serving cell and far away from all the neighboring cells so that the CIRs against these neighboring cells fall into the $Q$th CIR interval, resulting in that the CIRSP's elements corresponding to the 6 first-ring neighboring cells are $Q$ while the other elements are 0. Similarly, location B is close to the edge of the serving cell so that all the CIRs are low; cells 1,2,4,5,6 and 7 are the 6 strongest neighboring cells so that the corresponding CIRSP elements are 1,1, $Q1$, $Q1$, $Q1$ and $Q2$, and the other CIRSP elements are zero.

For the purpose of regression analysis in the following subsection,
the CIRSP is transformed into a spectrum profile (SP) matrix \mbox{$ {\bf{S}} =
[{\bf{s}}_1 {\kern 1pt} {\bf{s}}_2  \ldots {\bf{s}}_m  \ldots
{\bf{s}}_M ]_{IQ \times M} $} whose format matches to that of MMRs data. Its
$m$th column \mbox{$ {\bf{s}}_m  = \left[ {{\bf{s}}_{1,m}^{\rm T}
{\kern 1pt} {\bf{s}}_{2,m}^{\rm T} {\kern 1pt}  \ldots
{\bf{s}}_{i,m}^{\rm T} {\kern 1pt}  \ldots {\bf{s}}_{I,m}^{\rm T} }
\right]_{IQ \times 1}^{\rm T} $} is called CIRSP vector, and is
transformed in the following way from the CIRSP derived from the $m$th DT record
. Vector \mbox{$ {\bf{s}}_{i,m}  = {\bf{0}} $} if the CIRSP element corresponding to neighboring
cell $i$ is zero. The $q$th element of vector ${\bf{s}}_{i,m}$, which is the element in the $\left[ {\left( {i - 1} \right)Q + q} \right]$th row and $m$th column of ${\bf{S}}$ and is denoted as $s_{i,q,m}$, is one while all the other elements of vector ${\bf{s}}_{i,m}$ are zero if the CIRSP element corresponding to neighboring cell $i$ is $q$. That is,   $s_{i,q,m}$ equals to 1 if the neighboring cell $i$ is one of the 6 strongest neighboring cells and $d_{i,m}$  is fall into CIR interval $q$, it equals to 0 otherwise. An example of \mbox{$ {\bf{S}} $} is shown
in Figure 6. "1" appears at positions (1,2), (2,1) and (3,$Q$) since
the original CIRSP elements corresponding to neighboring cell 1,2,3
and 4 are 2, 1, $Q$ and 0 respectively.

\vspace{0.1in}
\subsection{Clustering and Cluster Center}

The $M$ CIRSPs derived from DT data is clustered into $K$ clusters by clustering analysis [23] according to CIRSP similarity. Each cluster corresponds to a characteristic region (called region for short) in the serving cell. Correspondingly, DT records are classified into the $k$th set if their CIRSPs belong to the $k$th cluster, \mbox{${k = 1, \ldots ,K}$}.

Membership vector ${\bf{A}}$, size vector ${\bf{B}}$  and cluster center matrix ${\bf{C}}$ are obtained through clustering analysis. \mbox{${\bf{A}} = \left[ {a_1 {\kern 1pt} a_2 {\kern 1pt}  \ldots {\kern 1pt} a_m  \ldots a_M } \right]_{1 \times M}$} , and its element $a_m  = k$ if the $m$th DT record is classified into the $k$th set. \mbox{${\bf{B}} = \left[ {b_1 {\kern 1pt} b_2 {\kern 1pt}  \ldots b_k {\kern 1pt}  \ldots b_K } \right]_{1 \times K}$} and $b_k$ is the number of DT records in the $k$th set, $b_1  + {\kern 1pt} b_2  + {\kern 1pt}  \ldots  + b_K  = M$. Vector ${\bf{B}}$ describes the geographic distribution of DT data over all the $K$ regions. And cluster center matrix is expressed as
\begin{equation}
{\bf{C}} = \left[ {{\bf{c}}_1 {\kern 1pt} {\bf{c}}_2 {\kern 1pt}  \ldots {\bf{c}}_k {\kern 1pt}  \ldots {\bf{c}}_K } \right]_{IQ \times K}  = \left[ {\begin{array}{*{20}c}
   {{\bf{c}}_{1,1} } & {{\kern 1pt}  \ldots } & {{\bf{c}}_{1,k} } &  \ldots  & {{\bf{c}}_{1K} }  \\
   {{\kern 1pt}  \ldots } & {{\kern 1pt}  \ldots } & {{\kern 1pt}  \ldots } & {{\kern 1pt}  \ldots } & {{\kern 1pt}  \ldots }  \\
   {{\bf{c}}_{i,1} } & {{\kern 1pt}  \ldots } & {{\bf{c}}_{i,k} } & {{\kern 1pt}  \ldots } & {{\bf{c}}_{i,K} }  \\
   {{\kern 1pt}  \ldots } & {{\kern 1pt}  \ldots } & {{\kern 1pt}  \ldots } & {{\kern 1pt}  \ldots } & {{\kern 1pt}  \ldots }  \\
   {{\bf{c}}_{I,1} } & {{\kern 1pt}  \ldots } & {{\bf{c}}_{I,k} } & {{\kern 1pt}  \ldots } & {{\bf{c}}_{I,K} }  \\
\end{array}} \right]
\end{equation}
where column vector ${\bf{c}}_k {\kern 1pt}$ of length $IQ$ is the cluster center of the $k$th cluster, while ${\bf{c}}_{i,k}$ is a column vector of length $Q$,
\begin{equation}
{\bf{c}}_{i,k}  = \left[ {c_{i,1,k} {\kern 1pt} c_{i,2,k} {\kern 1pt}  \ldots c_{i,q,k} {\kern 1pt}  \ldots c_{i,Q,k} {\kern 1pt} } \right]^{\rm T}
\end{equation}
Its $q$th element $c_{i,q,k}$ is the matrix element at the $\left[ {\left( {i - 1} \right)Q + q} \right]$th row and the $k$th column of ${\bf{C}}$,
\begin{equation}
c_{i,q,k}  = \frac{1}{{b_k }}\sum\limits_{\left\{ {a_m \left| {a_m  = k} \right.} \right\}} {s_{i,q,m} }
\end{equation}

For the $k$th cluster and neighboring cell $i$, $c_{i,q,k}$ is the probability that CIRSP element equals to $q$, ${\bf{c}}_{i,k}$ is the probability distribution of the CIRSP elements over all the $Q$ CIR intervals. For the $k$th cluster, ${\bf{c}}_k {\kern 1pt}$ gives $I$ probability distributions of the CIRSP elements over all the $Q$ CIR intervals and corresponding to $I$ neighboring cells.

\vspace{0.1in}
\subsection{Regression and Traffic Distribution}

The purpose of the regression is to estimate the number of MMRs belonging to each cluster. Assume that there are $N$ neighboring cells whose signal strength measurements have been contained in both MMRs and DT records, $N \le I$ and $N \le J$. These $N$ neighboring cells are called common neighboring cells. Picking out the elements corresponding to the common neighboring cells from vector ${\bf{R}}$ and matrix ${\bf{C}}$, as the samples of the dependent variable and the explanatory variables for multiple linear regression model.

Without loss of generality, we assume that the first $N$ neighboring cells of both MMRs data and DT data are the common neighboring cells, with index $n = 1 \ldots N$. Therefore there are $I-N$ neighboring cells whose signal strength measurements are contained only in DT records, with cell index $n' = N + 1, \ldots ,I$.

An augmented matrix is constructed as
\begin{equation}
{\bf{\bar C}} = \left[ {\begin{array}{*{20}c}
   {\bf{I}} & {\bf{C}}  \\
\end{array}} \right]_{IQ \times (K + 1)}
\end{equation}
where ${\bf{I}}$ is an all "1" vector of length $IQ$. Another augmented matrix ${\bf{\bar C'}}$ is constructed as
\begin{equation}
{\bf{\bar C'}} = \left[ {\begin{array}{*{20}c}
   {{\bf{I'}}} & {{\bf{C'}}}  \\
\end{array}} \right]_{NQ \times (K + 1)}
\end{equation}
where ${\bf{I'}}$ is an all "1" vector of length $NQ$, ${\bf{C'}}$ contains the top $NQ$ rows of ${\bf{C}}$ and can be expressed as
\begin{equation}
{\bf{C'}} = \left[ {{\bf{c'}}_1 {\kern 1pt} \;{\bf{c'}}_2 {\kern 1pt}  \ldots {\bf{c'}}_k {\kern 1pt}  \ldots {\bf{c'}}_K } \right]_{NQ \times K}
\end{equation}
thus vector ${\bf{c'}}_k$ contains the top $NQ$ rows of ${\bf{c}}_k$. Denote random error vector as
\begin{equation}
{\bm {\varepsilonup }} = \left[ {\varepsilon _1 {\kern 1pt} \varepsilon _2  \ldots \varepsilon _{NQ} } \right]^{\rm T}
\end{equation}
where $\varepsilon _1 ,{\kern 1pt} \varepsilon _2 , \ldots ,\varepsilon _{NQ}$ are i.i.d and obey normal distribution. Therefore, the multiple linear regression model can be formulated as
\begin{equation}
{\bf{R'}} = {\bf{\bar C'\bm \betaup } } + {\bm {\varepsilonup }}
\end{equation}
where ${\bf{R'}}$ is a column vector containing only the top $NQ$
rows of ${\bf{R}}$.

Regression model (8) stands because MMRs data is actually a linear combination of the \mbox{CIRSPs}
derived from DT records, if not considering the error resulted from the different measurement methods and tools. Therefore, MMRs data can be viewed approximately as a linear
combination of the $K$ cluster centers. The coefficients can be
calculated from (8) and compose a $\left( {K + 1} \right)$-dimensional vector
\begin{equation}
{\bm{\betaup }} = \left[ {\beta _0  \ldots \beta _k  \ldots \beta _K } \right]^{\rm{T}}
\end{equation}
where $\beta _k$ is actually the estimate of the number of the MMRs reported from the $k$th region. The constant $\beta _0$ is the estimate of the number of MMRs whose CIRSPs not included in DT records. $\beta _0$ is generally not zero since drive test might not go through the whole serving cell and thus lead to partially mismatch between the CIRSPs of MMRs and DT records. Apparently, ${\bm{\betaup }}$ is an estimate of practical traffic distribution over different regions.

\vspace{0.1in}
\subsection{Data Fusion and IM Generation}
Data fusion is done in the following way: the severe interfering neighboring cells omitted from MMRs are found and recovered from the DT data, and then complemented into MMRs data. A neighboring cell is a severe interfering neighboring cell if
\begin{equation}
\sum\limits_{q = 1}^{Q_{{\rm{threshold}}} } {\sum\limits_{k = 1}^K {\beta _k c_{n',q,k} } }  > 0
\end{equation}
where $Q_{{\rm{threshold}}}$ is the preset CIR interval threshold as stated in Section II-C. If CIR value against a neighboring cell falls into intervals numbered from 1 to $Q_{{\rm{threshold}}}$, this neighboring cell is a potential interference source because its signal is strong enough. If (10) is satisfied, neighboring cell $n'$ is one of the 6 strongest neighboring cells, and the CIR against neighboring cell $n'$ is lower than the CIR threshold in one or more regions.

If neighboring cell $n'$ satisfies (10) and its signal is only detected by drive test, it has been omitted from MMRs for some reason and needs to be complemented to MMRs data.

Assume there are $N'$ ($N' \le I - N$) omitted severe interfering neighboring cells in total, then the corresponding MMRs data ${\bf{R''}}$ can be estimated by the regression model as
\begin{equation}
{\bf{R''}} = {\bf{\bar C''\bm \betaup }}
\end{equation}
where ${\bf{\bar C''}}$ is a matrix containing $N'Q$ rows of the augmented matrix ${\bf{\bar C}}$ , and these N'Q rows are related with the severe interfering neighboring cells which are omitted from MMRs.

The fused MMRs data is then constructed from MMRs data ${\bf{R}}$ and the completed MMRs data ${\bf{R''}}$ as
\begin{equation}
{\bf{\hat R}} = \left[ {\begin{array}{*{20}c}
   {\bf{R}}  \\
   {{\bf{R''}}}  \\
\end{array}} \right]
\end{equation}
Therefore ${\bf{\hat R}}$ is called reinforced MMRs data.

Finally, an IM is generated from ${\bf{\hat R}}$ by using the way in Section II-C. This IM is named as IM-MR' for short.

For simplicity, the IM generation algorithm proposed in this section is named as MMRs+DT algorithm. The fused source data is the reinforced MMRs data, which is derived by complementing MMRs data with the originally omitted severe interfering neighboring cells provided by DT data.

\vspace{0.2in}
\section{IM GENERATION ALGORITHM BASED ON DT DATA RESHAPED BY MMRS DATA}

IM can also be generated using DT data as stated in Section I. However, we found that the geographical distribution of DT data ${\bf{B}} = \left[ {b_1 {\kern 1pt} b_2 {\kern 1pt}  \ldots b_k {\kern 1pt}  \ldots b_K } \right]_{1 \times K}$  relies on the vehicle velocity and road situation, thus are useless to frequency planning/optimization. If the geographical distribution is replaced by the actual communication traffic distribution, then the generated IM will contain traffic information and result in more reasonable frequency planning/optimization.

The algorithm proposed in this section is named as DT+MMRs for short. The block diagram of this algorithm is shown in Figure 7. Data Fusion II is the only block different from that in Figure 4, which generates the reshaped DT data by changing DT data's distribution ${\bf{B}}$ to ${\bm {\betaup }}$. The MMRs data and DT data in Figure 7 play different roles from those in Figure 4. The output of the algorithm is named IM-DT' to reflect that it is an IM generated from the reshaped DT data.

Changing the distribution of DT data from ${\bf{B}}$ to ${\bm{\betaup }}$ is achieved by replicating $b_k$ DT records to $\left\lfloor {\beta _k } \right\rfloor$ DT records for every set, where $\left\lfloor x \right\rfloor$   means the maximal integer no larger than $x$. The reshaping procedure for the $k$th set is as follows: each DT record of the set is replicated $E_k  = \left\lfloor {{{\beta _k } \mathord{\left/ {\vphantom {{\beta _k } {b_k }}} \right.
 \kern-\nulldelimiterspace} {b_k }}} \right\rfloor$ times, then $F_k  = \beta _k \bmod (b_k )$ DT records are picked out from the set randomly, and thus totally $E_k b_k  + F_k  \approx \beta _k$ DT records are obtained.

\vspace{0.2in}
\section{SIMULATION RESULTS}

Simulations were carried out to evaluate the effects of the proposed data fusions, and to make comparison between the IM generated from the fused data using our algorithms and that generated from the original data using traditional IM generation algorithms.

MMRs and DT records were collected from a
practical cellular system. The simulated area is shown in Figure 8
where base stations (BSs) are marked as five-pointed stars and the track
of the drive test is marked as dots. Cell 1 in
the top left of the figure is the serving cell. The
simulation settings are given in Table 1.

\vspace{0.1in}
\subsection{Clustering and Characteristic Regions}

The service signal of Cell 1 mainly covers two roads perpendicular to each other and the buildings along them. The drive test is executed along these two roads. All the obtained DT records are plotted in Figure 9 according to the longitudes and latitudes of the locations where the DT records are measured. By clustering analysis, the DT data of Cell 1 was clustered into 8 clusters, and the DT records were classified into 8 sets. DT records belonging to different sets are plotted with different markers in Figure 9. It can be observed from Figure 9 that clusters and locations are strongly related since the DT records obtained from a geographical region with unique wireless environment have the same CIRSP.

\vspace{0.1in}
\subsection{Regression and Traffic Distribution}

The explanatory variables ${\bf{c'}}_k {\kern 1pt}$ (${k = 1,2,
\ldots ,8}$) enter the regression model in turn according to the
order ranked by the partial correlation coefficient between
${\bf{c'}}_k {\kern 1pt}$ and dependent variable ${\bf{R'}}$. In
this simulated regression model, explanatory variables entered in the order of
${\bf{c'}}_2 {\kern 1pt}$, ${\bf{c'}}_1 {\kern 1pt}$, ${\bf{c'}}_3
{\kern 1pt}$  and ${\bf{c'}}_6 {\kern 1pt}$ . The significance
probability of the regression model is 0.0001 which is smaller than
the preset significance level (0.05), indicating that the dependent
variable is strongly related with all the explanatory variables
entered the model. The coefficient of determination is 0.858,
meaning that the well-done regression made MMRs data effectively
decomposed by 8 cluster centers. The model coefficients are listed
in Table 2, and they all passed the significance test.

\vspace{0.1in}
\subsection{Data Fusion and Obtained IMs}

Data Fusion I --- reinforcing MMRs data with originally omitted severe interfering neighboring cells. As to Cell 1, the solid five-pointed stars in Figure 8 represent the BS sites of the neighboring cells with the interfering frequencies. After applying Data Fusion I to Cell 1, three severe interfering neighboring cells omitted originally are recovered and their BS sites are marked as hollow five-pointed star in Figure 8. MMRs data of these three neighboring cells is calculated using (11) £¬and combined with the original MMRs data to form the reinforced MMRs data using (12). The obtained IM-MR' is surely different from the IM-MR obtained from the original data using traditional algorithms.

Data Fusion II --- reshaping the distribution of DT data. The reshaped DT data is obtained by DT+MMRs algorithm. The cumulative probability distributions of the original DT data and the reshaped DT data are illustrated in Figure 10 respectively. It can be found that the elements of IM generated from the reshaped DT data (marked as DT' data in Figure 10) is different from those of IM generated from DT data.

It is expectable that the similarity of IM-DT' and IM-MR is higher than that of IM-DT and IM-MR, since IM-DT' has traffic information and thus is more close to IM-MR. For the cells in Figure 8, calculating the correlation between IM-DT' and IM-MR, and that between IM-DT and IM-MR, the obtained correlation coefficients are 0.608 and 0.549 respectively. This verifies that the proposed Data Fusion II does provide useful traffic information.

IM generated from fused data. The IM of the simulation area is generated as IM-MR' and IM-DT' from reinforced MMRs data and reshaped DT data respectively. The Pearson correlation coefficient between IM-MR' and IM-DT' is 0.613, therefore these two IMs are highly correlated. It is reasonable because they are both the descriptions of the same wireless environment.

\vspace{0.2in}
\section{CONCLUSIONS}

DT records and MMRs describe the intra-system interference from the views of geographical coverage and customer experience respectively. Both of them have advantages and disadvantages. They can be fused to obtain an accurate IM with complete information.

In this paper, two IM generation algorithms based on source data fusion are proposed. The MMRs+DT algorithm is executed mainly based on MMRs data while DT data is used to reinforce the frequency-domain information of MMRs data. On the contrary, the DT+MMRs algorithm is mainly based on DT data while MMRs data are used to provide traffic distribution. IM-MR' and IM-DT' generated from the fused source data are more accurate theoretically than IM-MR and IM-DT generated from the original source data, respectively.

The simulation results show that the CIRSP is the "fingerprint" of position as the clusters match with the regions, MMRs data and DT data can be classified and matched according to CIRSP. DT data from a region can be used to recover the severe interfering neighboring cells omitted in the MMRs data from the same region. Geographical distribution of DT data over regions can be reshaped to practical traffic distribution. IM-MR' and IM-DT' are highly correlated so that each of them can be used for better frequency planning/optimization.

Since the source data is easy to obtained in conventional ways and no more assistant resources such as 3-dimensional digital map are needed, the proposed algorithms are applicable for practical engineering.

The proposed technique also provides a reference to 3G and future systems' frequency planning/optimization.

\vspace{0.2in}
\section*{ACKNOWLEDGMENTS}

This work was jointly supported by National Science and Technology Major Project of the Ministry of Science and Technology of China (2010ZX03001-001), Nature Science Foundation of China (60972058) and China Mobile Group Zhejiang Company Limited.

\newpage

\begin{figure}[!t]
\centering
\includegraphics[width=0.65\textwidth]{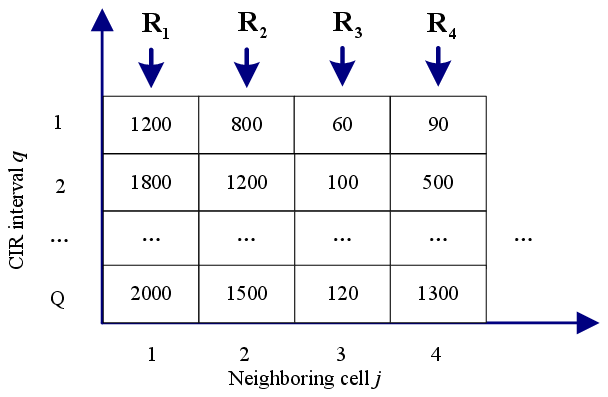}
\caption{An example of MMRs data.} \label{fig1}
\end{figure}

\begin{figure}[!t]
\centering
\includegraphics[width=0.8\textwidth]{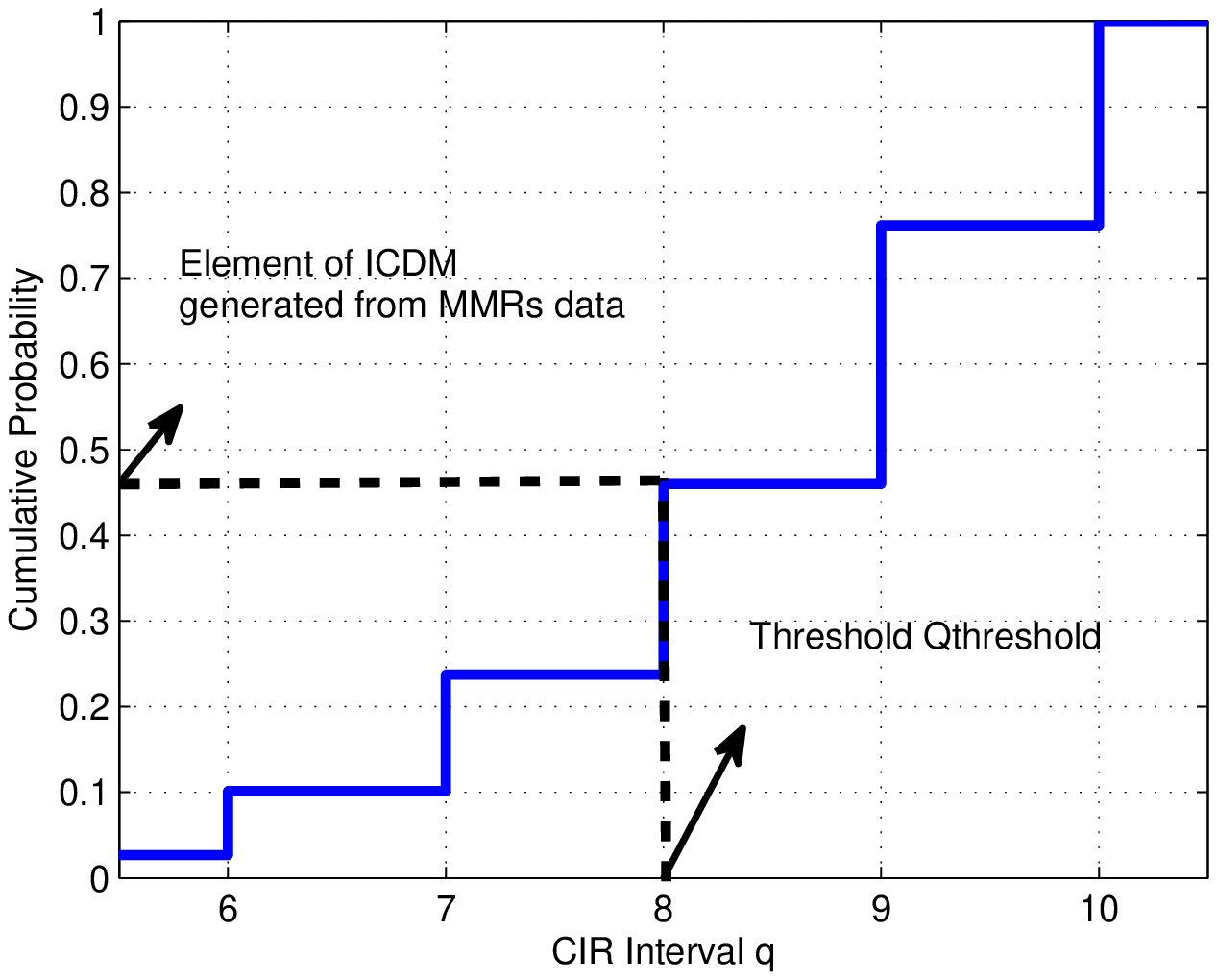}
\caption{The procedure of obtaining ICDM element generated from MMRs data.} \label{fig2}
\end{figure}

\begin{figure}[!t]
\centering
\includegraphics[width=0.8\textwidth]{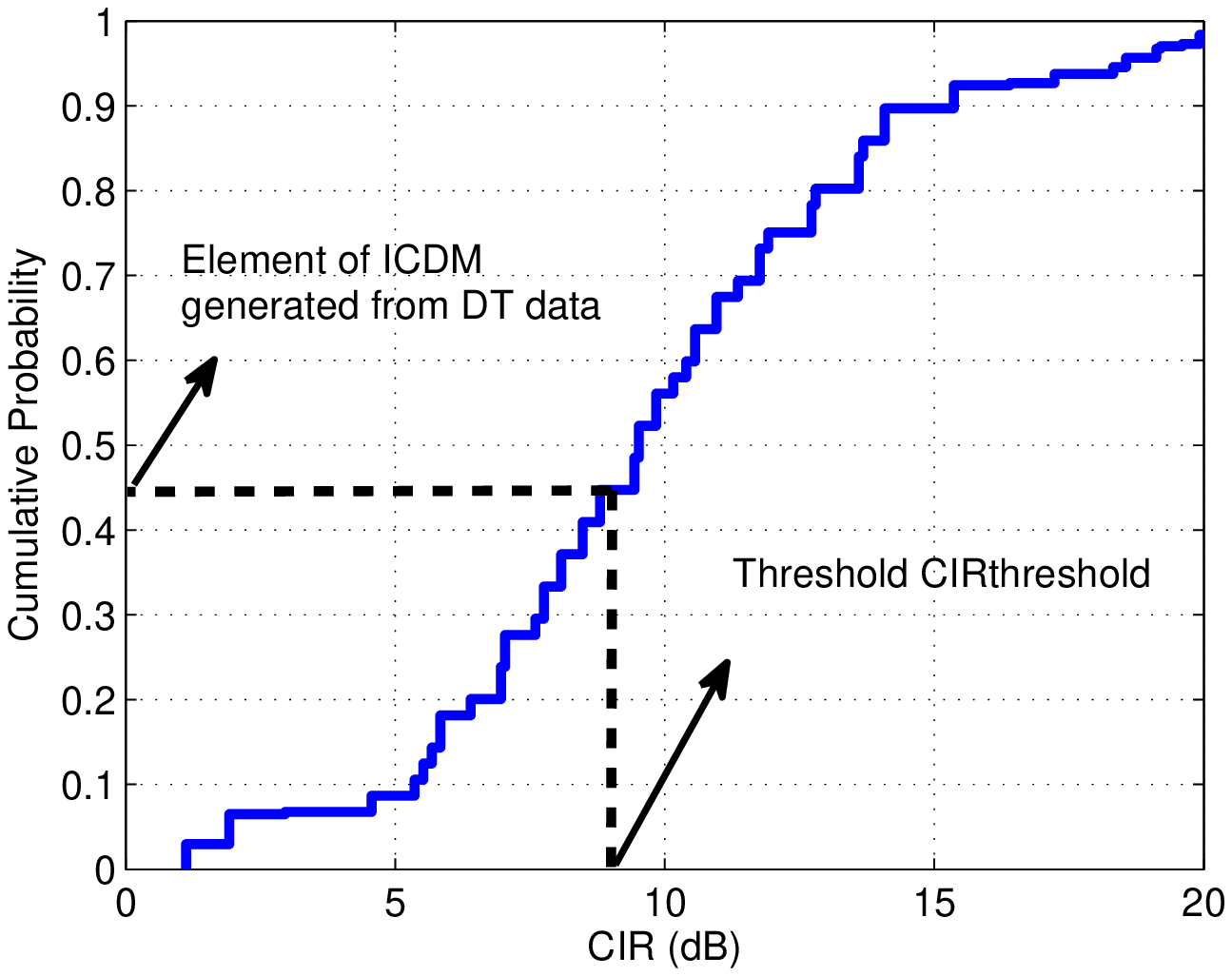}
\caption{The procedure of obtaining ICDM element generated from DT data.}
\label{fig3}
\end{figure}

\begin{figure}[!t]
\centering
\includegraphics[width=0.8\textwidth]{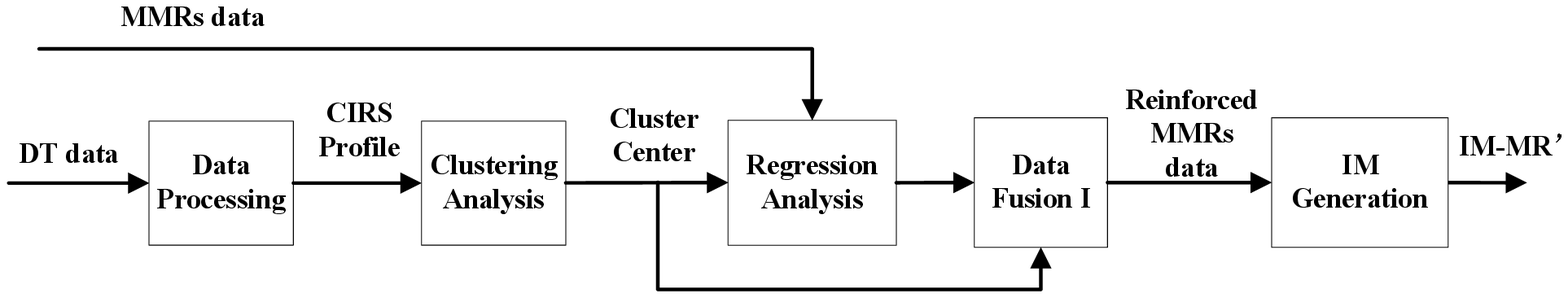}
\caption{ The procedure of MMRs+DT algorithm.}
\label{fig4}
\end{figure}

\begin{figure}[!t]
\centering
\includegraphics[width=0.8\textwidth]{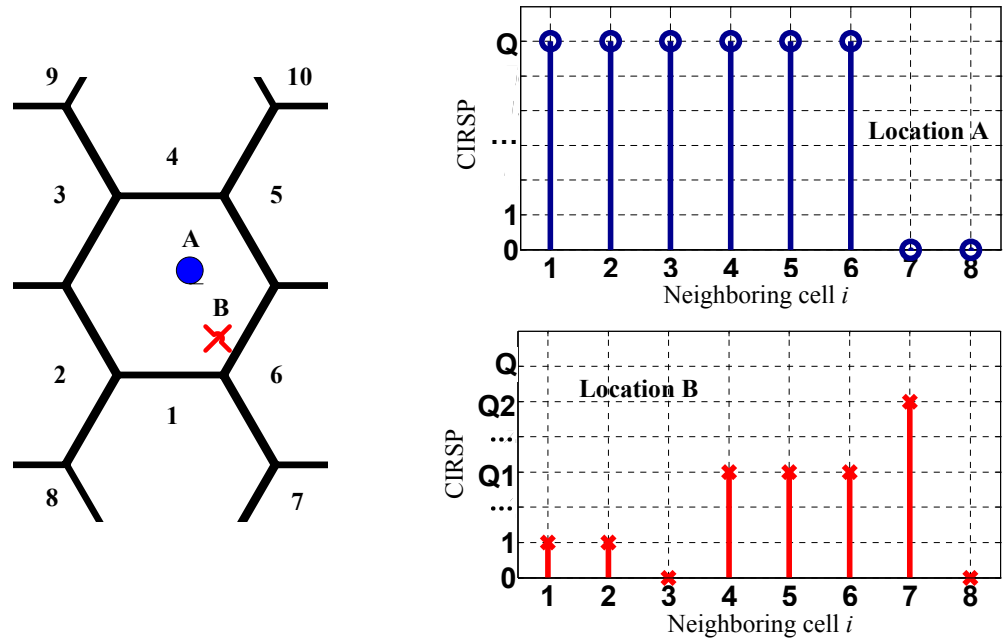}
\caption{Measurement positions and CIRSPs.} \label{fig5}
\end{figure}

\begin{figure}[!t]
\centering
\includegraphics[width=0.7\textwidth]{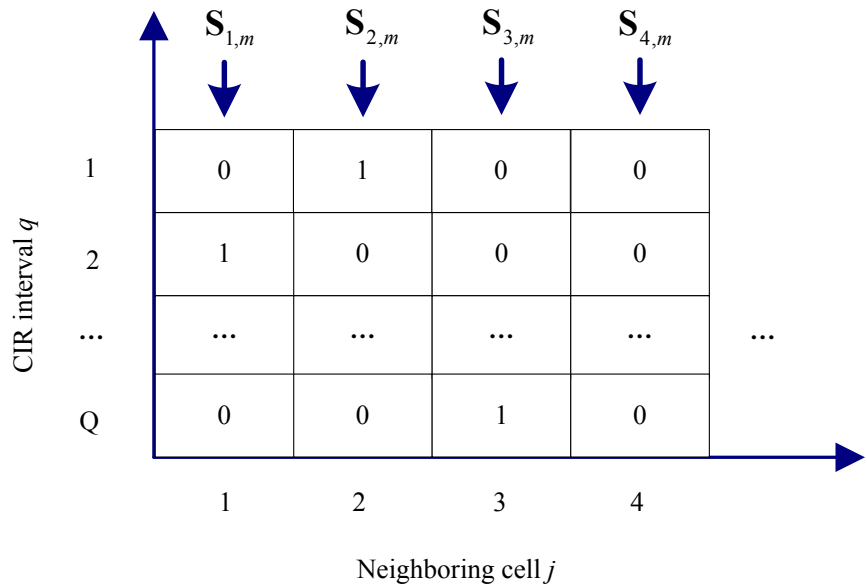}
\caption{SP matrix ${\bf{S}}$, CIRSP vector ${\bf{s}_{i,m}}$ and their elements $s_{i,q,m}$.} \label{fig6}
\end{figure}

\begin{figure}[!t]
\centering
\includegraphics[width=0.8\textwidth]{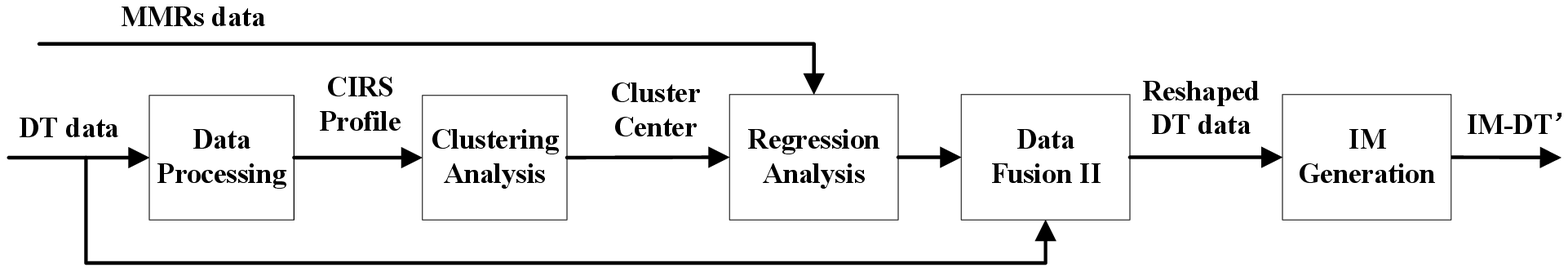}
\caption{ The procedure of DT+MMRs algorithm.}
\label{fig7}
\end{figure}

\begin{figure}[!t]
\centering
\includegraphics[width=0.8\textwidth]{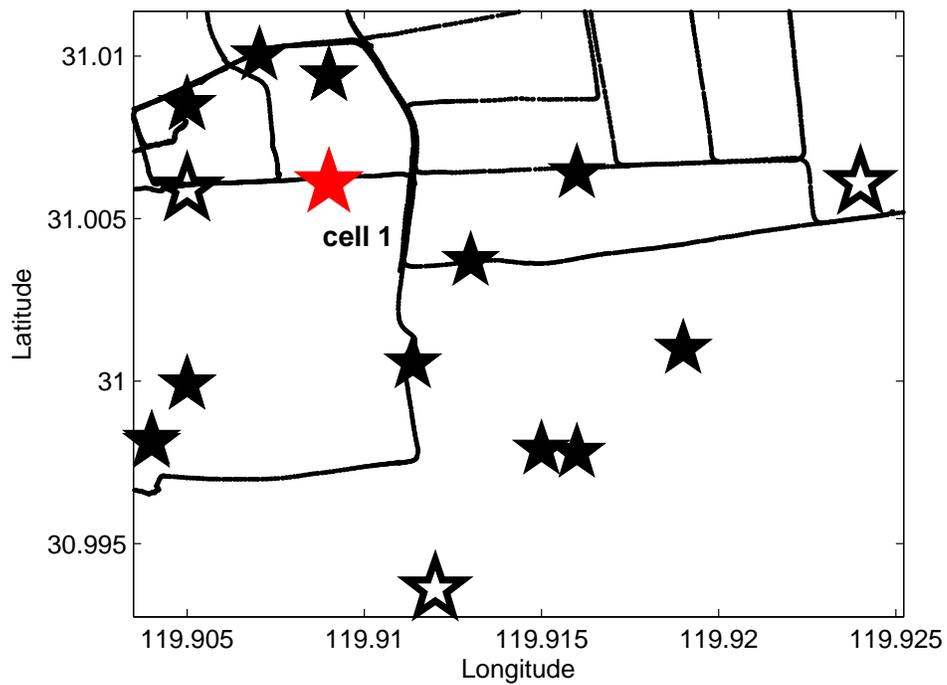}
\caption{Geographical positions of the cells in the simulated area.}
\label{fig8}
\end{figure}

\begin{table}[!t] \caption{\label{tab:test}Simulation settings}
\begin{center}
 \begin{tabular}{|c|c|}
  \hline
  DT data (cell 1) & $M$=609, $I$=123 \\
  \hline
  MMRs data (cell 1) & $Q$=10, $J$=103\\
  \hline
  clustering & K-means clustering with K=8\\
  \hline
   & Stepwise regression [23].\\
   regression & Significance level for variable to enter model is 0.05 and\\
   &  that for variable to be removed is 0.10.\\
  \hline
 \end{tabular}
\end{center}
\end{table}

\begin{figure}[!t]
\centering
\includegraphics[width=0.8\textwidth]{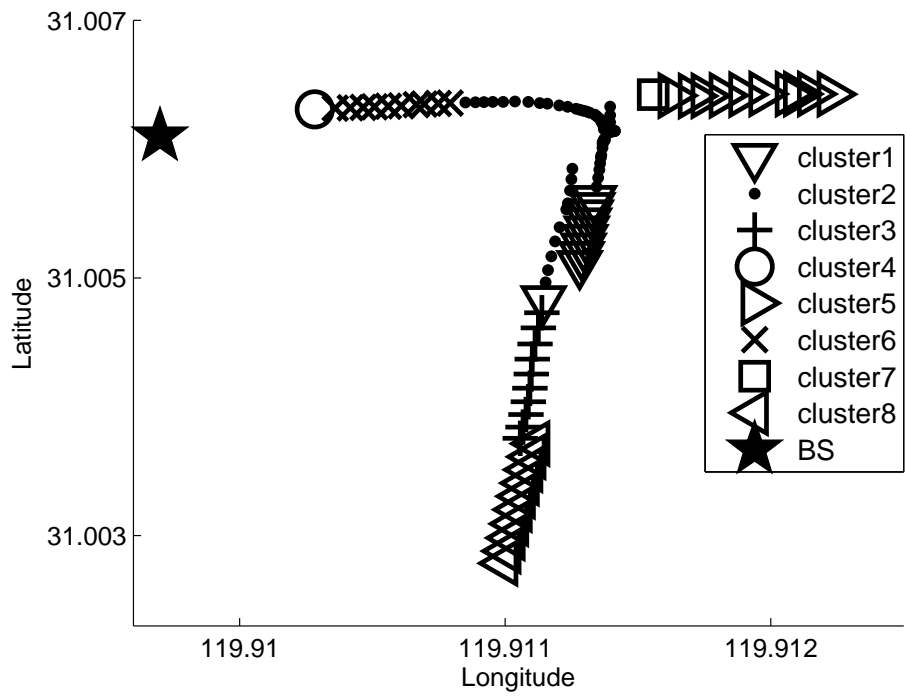}
\caption{8 sets of DT records belong to 8 regions in serving cell respectively.}
\label{fig9}
\end{figure}

\begin{table}[!t] \caption{\label{tab:test}The coefficients of regression}
\begin{center}
 \begin{tabular}{|c|c|c|c|c|}
  \hline
  $\beta _0$ & $\beta _2$ & $\beta _1$ & $\beta _3$ & $\beta _6$ \\
  \hline
  31391.563 & 1377763.572 & 1724748.834 & 842328.720 & 182019.175 \\
  \hline
 \end{tabular}
\end{center}
\end{table}

\begin{figure}[!t]
\centering
\includegraphics[width=0.8\textwidth]{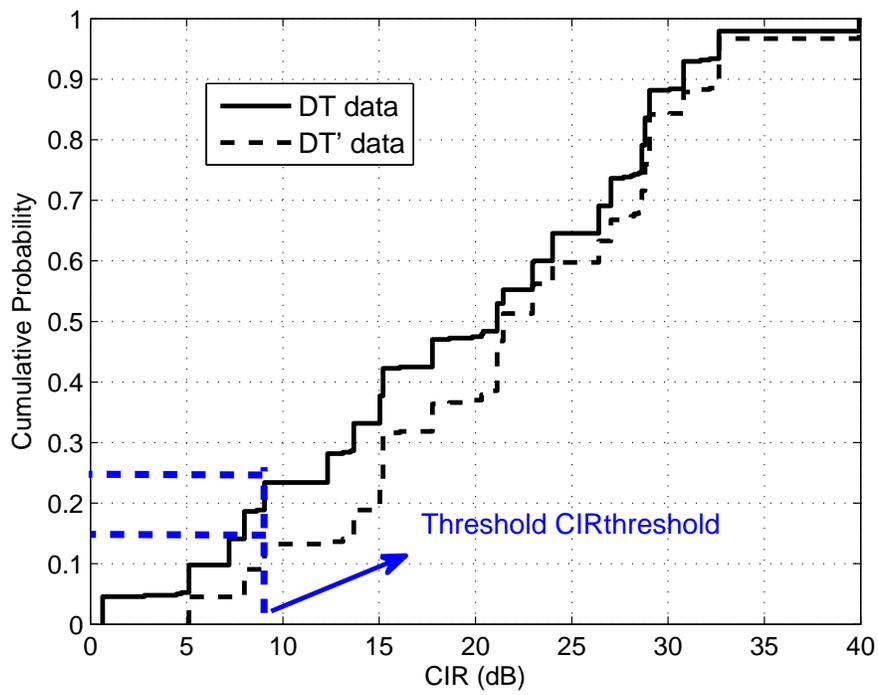}
\caption{The cumulative probability distributions of the CIR against
a neighboring cell.} \label{fig10}
\end{figure}

\end{document}